\documentclass[acmsmall,screen]{acmart}
\setcopyright{none}
\copyrightyear{2025}
\acmYear{2025}
\acmISBN{}
\acmDOI{}

\usepackage{fancyhdr}

\usepackage{amsmath,subcaption,bussproofs}
\usepackage{minted}[frozencache]

\newtheorem{lemma}{Lemma}[section]
\newtheorem{coqlemma}[lemma]{Lemma$^\star$}

\title{An Empirical Study of Rational Tree Unification for \textsc{miniKanren}}

\author{Eridan Domoratskiy}
\email{eridan200@mail.ru}
\orcid{0009-0005-9020-3403}

\author{Dmitrii Kosarev}
\email{d.kosarev@spbu.ru}
\email{kosarev.d.s@yandex.ru}
\orcid{0000-0002-6773-5322}

\author{Dmitry Boulytchev}
\email{dboulytchev@math.spbu.ru}
\orcid{0000-0001-8363-7143}

\affiliation{%
  \institution{St.~Petersburg State University}
  \city{St.~Petersburg}  
  \country{Russia}
}



\begin{document}

\settopmatter{printacmref=false}
\settopmatter{printfolios=true}
\renewcommand\footnotetextcopyrightpermission[1]{}
\pagestyle{fancy}
\fancyfoot{}
\fancyfoot[R]{miniKanren'25}
\fancypagestyle{firstfancy}{
  \fancyhead{}
  \fancyhead[R]{miniKanren'25}
  \fancyfoot{}
}
\makeatletter
\let\@authorsaddresses\@empty
\makeatother

\begin{abstract}
We present a study of unification for rational trees in the context of \textsc{miniKanren}. 
We give the definition of rational trees, specify the unification algorithm and prove some
of its properties. We also introduce a number of heuristic optimizations and evaluate them
for a number of relevant benchmarks. Finally we discuss the relations between rational
and conventional unification algorithms and possible scenarios of their coexistence
in the context of relational programming.
\end{abstract}

\maketitle
\thispagestyle{firstfancy}
\section{Introduction}

The classical interpretation of logic programs is the least Herbrand model which consists of finite trees made of variables and constructors, and the conventional understanding of relational programs is aligned with this interpretation~\cite{rozplokhas2019certified}.

The only predicate used in defining the least Herbrand model is a syntactical equality which in logic programs is handled by unification; in particular in \textsc{miniKanren} a classical Robinson's algorithm over a triangular substitution with ``occurs check'' is used as a rule. Since the least Herbrand model operates with finite trees only, ``occurs check'' is needed to prohibit recursively infinite trees. 

However, in many \textsc{Prolog} implementations ``occurs check'' is suppressed due to performance considerations. For example, list concatenation implemented in \textsc{Prolog} with ``occurs check'' takes a quadratic time instead of expected linear time~\cite{colmerauer1982prolog}. This simplification makes unification generally unsound w.~r.~t. the least Herbrand model since it does not deny recursive trees.

To address this issue, Colmerauer presented the rational tree model~\cite{colmerauer1982prolog}. Rational tree (or ``regular tree'') is a possibly infinite tree with finite cardinality of its subtree set. In other words, rational trees could be infinite but only by recursively repeating some of their finite parts. While this interpretation admits recursive trees ``occurs check'' cannot be simply removed since with conventional unification algorithms this would still results in an infinite looping. Thus, the whole unification algorithm has to be revisited.

There are several practical rational tree unification algorithms presented before, one of the latest is ``UNIFY-1'' by Martelli et.~al.~\cite{martelli1984efficient}. These algorithms replace substitution with system of equations of different flavors and iteratively perform simplification steps.

Besides the performance trade-offs, rational tree unification is notable for dealing with naturally recursive structures like equirecursive types. For example, there was a previous attempt to implement a relational solver for type inference with equirecursive types using \textsc{OCanren} with classical unification with ``occurs check''~\cite{domoratskiy2024relational}. The authors faced an issue with the minimality of produced answers, e.~g. their solver produced the type $Nil \sqcup Cons(\mathbb{Z}, \mu a. ~ Nil \sqcup Cons(\mathbb{Z}, a))$ instead of a simpler one $\mu a. ~ Nil \sqcup Cons(\mathbb{Z}, a)$. The minimality problem for finite representation of rational trees was addressed before by Courcelle et.~al.~\cite{courcelle1974algorithms} but for efficient implementation their algorithm requires an equation system in a special form.

In this paper we present a novel unification algorithm based on that of~\cite{martelli1984efficient} but simpler in implementation and making it possible to perform minimization of the produced equation system as well. We prove the termination and unifier properties for the algorithm and discuss its other properties. Presented algorithm is implemented in place of that currently used in \textsc{OCanren} and evaluated on a number of examples including equirecursive type inference. Possible variants of implementation including those with ``occurs check'' are studied, too.

In the rest of the text we use the term ``tree'' instead of a more conventional term ``term'' in order to be more coherent with ``rational trees'' terminology.

Some of the lemmas presented in the paper are supported by corresponding \textsc{Coq} proofs; those lemmas are marked with ``$\star$''.

\section{Rational Trees}

In this section we introduce rational trees, the syntax for their finite representation, and discuss some of their relevant properties.

First, we define the set of variables ($\mathbb{V}$) and the set of constructors ($\mathbb{C}$) as countably infinite sets. Also, we assume there is a function $arity : \mathbb{C} \rightarrow \mathbb{N}$ which defines the arities of constructors and denote constructors as $f^n \in \mathbb{C}$, where $f \in \mathbb{C}$ is a constructor and $n = arity(f)$ is its arity.

We define the set of finite trees ($T$) as the least fixed point of the following syntactic definition:

\[T ::= \mathbb{V} \mid \mathbb{C}^n(T_1, ..., T_n).\]

On the other hand, we define the set of (possibly) infinite trees ($T^\infty$) as the \textit{greatest} fixed point of the same syntactic definition. As a result, $T$ consists of all finite Herbrand trees and $T_\infty$ consists of all possibly infinite Herbrand trees, thus $T \subseteq T^\infty$. 

We define the set of all subtrees of a given possibly infinite tree $t \in T^\infty$ in a natural way:

\begin{figure}[H]
    \centering
    \begin{subfigure}[B]{0.3\textwidth}
        \centering
        \AxiomC{}
        \UnaryInfC{$t \in subtrees(t)$}
        \DisplayProof
    \end{subfigure}
    \begin{subfigure}[B]{0.3\textwidth}
        \centering
        \AxiomC{$t \in subtrees(t_i)$}
        \UnaryInfC{$t \in subtrees(f^n(..., t_i, ...))$}
        \DisplayProof
    \end{subfigure}
\end{figure}

Now we can define the set of all rational trees ($T^R$) as the set of all possibly infinite trees that have only a finite number of subtrees:
\[T^R = \{ t \in T^\infty \mid subtrees(t) \text{ is finite}\}.\]

%We say that the set of subtrees is finite iff it consists of a finite number of different trees. 

For example, $f(f(f(...)))$ is a rational tree since it has a single subtree (itself) but $f_1(f_2(f_3(...)))$ where the ellipsis stands for the sequence of different constructor applications is not a rational tree.

Finally, we define the set of all $\mu$-trees ($T^\mu$) as the least fixed point of the following syntactic definition:
\[T^\mu ::= \mathbb{V} \mid \mathbb{C}^n(T^\mu_1, ..., T^\mu_n) \mid \mu \mathbb{V}. ~ T^\mu,\]
where $\mu$-binder is used to bind a variable to the tree itself to refer to it recursively. For example, $\mu x. ~ f(x)$ represents the rational tree $f(f(f(...)))$. Obviously, $T \subseteq T^\mu$.

We define variable substitution in a conventional way and denote the substitution of all free occurrences of a variable $x$ in a tree $t$ with a tree $t'$ as $t [x \mapsto t']$. The substitution naturally extends for possibly infinite trees and $\mu$-trees with conventional $\alpha$-conversion to prevent free variable capturing.

We define an ``unfolding step'' ($\rightarrow_\mu$) as the minimal binary relation that holds for all trees such that, first,

\[\mu x. ~ t \rightarrow_\mu t [x \mapsto \mu x. ~ t]\]

and, second, it is homomorphically extended to all trees in the left hand side, so that it performs a single unfolding step in all top-level occurrences of $\mu$-binders. For example, $\mu x. ~ f(x) \rightarrow_\mu f(\mu x. ~ f(x))$ and $g(\mu x. ~ f(g(x))) \rightarrow_\mu g(f(g(\mu x. ~ f(g(x)))))$. ``Top-level'' here means that we do not unfold $\mu$-binders inside other $\mu$-binders. Finally, we define the ``unfold'' relation ($\twoheadrightarrow_\mu$) as a transitive closure limit of unfolding steps. 

Our expectation is that the unfolding would always lead to an infinite rational tree:

\[\mu x. ~ f(x) \rightarrow_\mu f(\mu x. ~ f(x)) \rightarrow_\mu f(f(\mu x. ~ f(x))) \rightarrow_\mu ... \rightarrow_\mu f(f(f(...))).\]

However, there are some cases when it does not. For example, a $\mu$-tree $\mu x. ~ x$ unfolds to itself:

\[\mu x. ~ x \rightarrow_\mu \mu x. ~ x \rightarrow_\mu ... \rightarrow_\mu \mu x. ~ x,\]
and thus does not correspond to any rational tree. On the other hand, $\mu x. ~ f() \rightarrow_\mu f()$ is a finite rational.

%We call these $\mu$-trees not well-formed since they do not have an interpretation as finite representations of any rational tree.

We define a well-formedness property $wf(t)$ on $\mu$-trees inductively:
\begin{figure}[H]
    \centering
    \begin{subfigure}[B]{0.2\textwidth}
        \centering
        \AxiomC{}
        \RightLabel{(VarWF)}
        \UnaryInfC{$wf(x)$}
        \DisplayProof
    \end{subfigure}
    \begin{subfigure}[B]{0.35\textwidth}
        \centering
        \AxiomC{$\forall i, ~ wf(t_i)$}
        \RightLabel{(ConsWF)}
        \UnaryInfC{$wf(f^n(t_1, ..., t_n))$}
        \DisplayProof
    \end{subfigure}
    \begin{subfigure}[B]{0.35\textwidth}
        \centering
        \AxiomC{$wf(f^n(t_1, ..., t_n))$}
        \RightLabel{(BindWF)}
        \UnaryInfC{$wf(\mu x. ~ f^n(t_1, ..., t_n))$}
        \DisplayProof
    \end{subfigure}
\end{figure}

%The strong well-formedness property is slightly less permissive than the well-formedness property given before but doesn't reduce expressivity of well-formed $\mu$-trees. To prove this, we firstly define $\mu$-trees equivalence as equality of their unfoldings. Now we may see that we forbid two general cases of well-formed $\mu$-trees:
%\begin{itemize}
%    \item $\mu x. ~ y$ when $x \neq y$ --- after the first unfolding step turns into just $y$,
%    \item $\mu x_1. ~ \mu x_2. ~ ... ~ \mu x_n. ~ t$ --- all of $x_i$ will refer to equivalent $\mu$-trees.
%\end{itemize}

%Hence, we may state that unfolding is a total function from the set of well-formed $\mu$-trees to the set of infinite trees. Further we will mention well-formed $\mu$-trees as just ``$\mu$-trees'' in the most cases.

\begin{coqlemma}
    Unfolding of a well-formed $\mu$-tree is always a rational tree.
\end{coqlemma}
\begin{proof}
    Given a well-formed $\mu$-tree, consider it's unfolding. Every subtree of the unfolding is an unfolding of some subtree of the initial $\mu$-tree, so the cardinality of the set of the unfolding subtrees cannot be greater that the cardinality of the set of the initial $\mu$-tree subtrees.
\end{proof}

As a corollary, we may state that unfolding maps well-formed $\mu$-trees to the set of rational trees. Although it is obvious that this function is not injective (because we may encode a single rational tree in several ways), its surjectivity holds~--- there is a $\mu$-tree representation for any rational tree.

\begin{lemma}
    Unfolding is surjective.
\end{lemma}
\begin{proof}
    Well-founded induction on the number of subtrees of the given rational tree. In the base case we have exactly one subtree so it must be a variable or a constant (a constructor without arguments).
    
    In the induction step we have at most $n + 1$ subtrees. Let us consider two cases:
    \begin{itemize}
        \item If the current subtree is not a subtree of any of its subtrees then they all must have strictly fewer number of subtrees than current one so we may use the induction hypothesis directly.
        
        \item Otherwise, pick a fresh variable $x$ and substitute every occurrence of the current tree in subtrees with $x$. After that they all must have strictly fewer subtrees so we may build a $\mu$-tree representations for them. Now we may put them under the current constructor and wrap it in $\mu$-binder over $x$.
    \end{itemize}
    
    It's easy to see that the unfolding of the constructed $\mu$-tree is equivalent to the original rational tree. Moreover, the well-formedness property holds in all cases.
\end{proof}

\section{Rational Tree Unification}

In this section we define some data structures used in the unification algorithm, the algorithm itself with some key ideas concerning it's termination and correctness, and discuss some implementation details.

\subsection{Equation System}

We define \textit{head tree} $T^H$ as a constructor applied to variables:
\[T^H ::= \mathbb{C}^n(\mathbb{V}_1, ..., \mathbb{V}_n).\]
For example, $f(x, y, z)$ is a head tree but $f(g(x), y)$ is not --- it's a \textit{complex tree}.

Now we can define an equation $\mathcal{E}$ to be a triple of non-empty set of variables (denoted by $\mathbb{V}_+$), a \textit{representing} variable and an optional head tree (denoted by $T^H_?$):
\[\mathcal{E} ::= \mathbb{V}_+ \equiv \mathbb{V}, T^H_?.\]

Additionally we require that representing variable is  included in the set of variables in the left hand side of an equation. Missing head tree in the right hand side of an equation is denoted by $\bot$.

Finally, we define an equation system ($\Xi$) to be a finite set of equations with pairwise disjoint left hand sides. The domain $dom(\Xi)$ of equation system is the union of its left hand sides. The image $image_\Xi : \mathbb{V} \rightarrow T^\infty$ of equation system is defined by the following inference rules which are treated coinductively:

\begin{figure}[H]
    \centering
    \begin{subfigure}[B]{0.35\textwidth}
        \centering
        \AxiomC{$x \not\in dom(\Xi)$}
        \RightLabel{(NoImage)}
        \UnaryInfC{$image_\Xi(x) = x$}
        \DisplayProof
    \end{subfigure}
    \begin{subfigure}[B]{0.5\textwidth}
        \centering
        \AxiomC{$[X \equiv y, \bot] \in \Xi$}
        \AxiomC{$x \in X$}
        \RightLabel{(NoTreeImage)}
        \BinaryInfC{$image_\Xi(x) = y$}
        \DisplayProof
    \end{subfigure}

    ~

    \begin{subfigure}[B]{0.9\textwidth}
        \centering
        \AxiomC{$[X \equiv y, f^n(x_1, ..., x_n)] \in \Xi$}
        \AxiomC{$x \in X$}
        \RightLabel{(TreeImage)}
        \BinaryInfC{$image_\Xi(x) = f^n(image_\Xi(x_1), ..., image_\Xi(x_n))$}
        \DisplayProof
    \end{subfigure}
\end{figure}

%This definition is coinductive since we guarantee to produce at least one part of a resulting infinite tree and an inference tree may be infinite in case of recursively mentioned variables at the right hand sides of occurred equations. 

The image of an equation system allows us to treat it as a variable substitution to the set of infinite trees. It's easy to see that the resulting trees are always rational trees since we have only a finite number of equations in the system and all constructors have finite arities. 

We define a ``$\mu$-image'' as a recursive procedure which, given a variable and an equation system, returns a $\mu$-tree representing the image of this variable w.~r.~t. given equation system. In order to define it we need to describe an auxiliary data structure (``visited'') as a partial mapping from variables to booleans. In this mapping we need to distinguish between three states of a variable:

\begin{itemize}
    \item ``not encountered'' ($\bot$) --- we didn't encounter (yet) the variable on the path from the start variable to the given one;
    \item ``false'' ($F$) --- we encountered  given variable for the first time;
    \item ``true'' ($T$) --- we encountered given variable multiple times.
\end{itemize}

We define $\mu$-image by means of the following inference system:

\begin{figure}[H]
    \centering
    \begin{subfigure}[B]{0.45\textwidth}
        \centering
        \AxiomC{$x \not\in dom(\Xi)$}
        \RightLabel{(NoMuImage)}
        \UnaryInfC{$\mu\text{-}image_\Xi(x, vis) = x, vis$}
        \DisplayProof
    \end{subfigure}
    \begin{subfigure}[B]{0.5\textwidth}
        \centering
        \AxiomC{$[X \equiv y, \bot] \in \Xi$}
        \AxiomC{$x \in X$}
        \RightLabel{(NoTreeMuImage)}
        \BinaryInfC{$\mu\text{-}image_\Xi(x, vis) = y, vis$}
        \DisplayProof
    \end{subfigure}

    ~

    \begin{subfigure}[B]{0.9\textwidth}
        \centering
        \AxiomC{$[X \equiv y, f^n(x_1, ..., x_n)] \in \Xi$}
        \AxiomC{$x \in X$}
        \AxiomC{$vis(y) \neq \bot$}
        \RightLabel{(VarMuImage)}
        \TrinaryInfC{$\mu\text{-}image_\Xi(x, vis) = y, (vis, y \mapsto T)$}
        \DisplayProof
    \end{subfigure}

    ~

    \begin{subfigure}[B]{0.95\textwidth}
        \centering
        \AxiomC{$[X \equiv y, f^n(x_1, ..., x_n)] \in \Xi$}
        \AxiomC{$x \in X$}
        \AxiomC{$vis(y) = \bot$}
        \AxiomC{$vis_0 = vis, y \mapsto F$}
        \noLine
        \QuaternaryInfC{$t_1, vis_1 = \mu\text{-}image_\Xi(x_1, vis_0)$ \quad $...$ \quad $t_n, vis_n = \mu\text{-}image_\Xi(x_n, vis_{n - 1})$ \quad \fbox{$vis_n(y) = F$}}
        \RightLabel{(TreeMuImage)}
        \UnaryInfC{$\mu\text{-}image_\Xi(x, vis) = f^n(t_1, ..., t_n), (vis_n, y \mapsto \bot)$}
        \DisplayProof
    \end{subfigure}

    ~

    \begin{subfigure}[B]{0.95\textwidth}
        \centering
        \AxiomC{$[X \equiv y, f^n(x_1, ..., x_n)] \in \Xi$}
        \AxiomC{$x \in X$}
        \AxiomC{$vis(y) = \bot$}
        \AxiomC{$vis_0 = vis, y \mapsto F$}
        \noLine
        \QuaternaryInfC{$t_1, vis_1 = \mu\text{-}image_\Xi(x_1, vis_0)$ \quad $...$ \quad $t_n, vis_n = \mu\text{-}image_\Xi(x_n, vis_{n - 1})$ \quad \fbox{$vis_n(y) = T$}}
        \RightLabel{(MuTreeMuImage)}
        \UnaryInfC{$\mu\text{-}image_\Xi(x, vis) = \mu y. ~ f^n(t_1, ..., t_n), (vis_n, y \mapsto \bot)$}
        \DisplayProof
    \end{subfigure}
\end{figure}

The notation \mbox{$vis, x \mapsto v$} stands for the extending mapping $vis$ with binding \mbox{$x \mapsto v$} so \break \mbox{$(vis, x \mapsto v)(x) = v$} and \mbox{$(vis, x \mapsto v)(y) = vis(y), y \neq x$} hold. Initially, ``visited'' mapping is assumed to be empty. Rule (VarMuImage) guarantees the termination of the algorithm since no variable is extracted infinitely. The last two rules differ in the last check for the variable $y$ in the final ``visited'' mapping: if the variable was encountered during the calculation of its own $\mu$-image we need to enclose it by a corresponding $\mu$-binder. It's easy to see that $\mu$-image is always well-formed.

\begin{lemma}
    Given equation system $\Xi$ and variable $x$ the unfolding of $\mu\text{-}image_\Xi(x)$ equals to $image_\Xi(x)$.
\end{lemma}
\begin{proof}
    The proof sketch:
    \begin{enumerate}
        \item consider a simplified algorithm $\mu\text{-}image'_\Xi : \mathbb{V} \rightarrow T^\mu$ that erases variable from $dom(\Xi)$ instead of transforming
        the equation system using ``visited'' mapping;
        % using ``visited'' getting new equation system $\Xi'$;
        \item note that in the simplified algorithm $vis_n(x) = T$ holds iff the resulting tree (consisting of $\mu\text{-}image'_{\Xi'}(x_i)$ where $[X \equiv x, f^n(..., x_i, ...)] \in \Xi$) has free occurrences of $x$;
        \item the correctness proof of $\mu\text{-}image'_\Xi$ goes by the induction on the size of $dom(\Xi)$.
    \end{enumerate}
\end{proof}

In our \textsc{OCanren}-based implementation we use the procedure of $\mu$-image calculation instead of classical substitution application. On the other hand, we require input trees to be ``normal'' trees with no $\mu$-binders as in conventional \textsc{miniKanren}. This doesn't reduce the expressivity since we always can encode recursive trees by several unifications but imposes some difficulties in the internal implementation, e.~g. for tabling which is not supported yet.

To deal with the equation system we define an abstract interface consisting of three basic operations:

\begin{enumerate}
    \item $\emph{find}_\Xi : \mathbb{V} \rightarrow \mathbb{V} \times T^H_?$. Finds an equation by the given left hand side variable and returns its right hand side.
    \item $bind_\Xi : \mathbb{V} \times T^H \rightarrow \Xi$. Binds given variable to the given head tree.
    \item $union_\Xi : \mathbb{V} \times \mathbb{V} \rightarrow \Xi \times (T^H \times T^H)_?$. Unions given pair of variables and optionally returns a pair of head trees bound to them before union if both existed.
\end{enumerate}

More formally, these operations are defined as follows. For convenience we use the notation $\Xi, \mathcal{E}$ instead of $\Xi \cup \{\mathcal{E}\}$ and assume that an equation being added to a system \emph{replaces} all equations in the system in which their left-hand side sets have  non-empty intersections with that of the given equation (so $\Xi, (X \equiv x, t)$ denotes $\Xi \setminus \{Y \equiv y, t'\}, (X \equiv x, t)$ when $X \cap Y \neq \varnothing$).
\begin{figure}[H]
    \centering
    \begin{subfigure}[B]{0.45\textwidth}
        \centering
        \AxiomC{$x \not\in dom(\Xi)$}
        \UnaryInfC{$\emph{find}_\Xi(x) = x, \bot$}
        \DisplayProof
    \end{subfigure}
    \begin{subfigure}[B]{0.5\textwidth}
        \centering
        \AxiomC{$[X \equiv x', t] \in \Xi$}
        \AxiomC{$x \in X$}
        \BinaryInfC{$\emph{find}_\Xi(x) = x', t$}
        \DisplayProof
    \end{subfigure}

    ~

    \begin{subfigure}[B]{0.45\textwidth}
        \centering
        \AxiomC{$x \not\in dom(\Xi)$}
        \UnaryInfC{$bind_\Xi(x, t) = \Xi, (\{x\} \equiv x, t)$}
        \DisplayProof
    \end{subfigure}
    \begin{subfigure}[B]{0.5\textwidth}
        \centering
        \AxiomC{$[X \equiv x', t'] \in \Xi$}
        \AxiomC{$x \in X$}
        \BinaryInfC{$bind_\Xi(x, t) = \Xi, (X \equiv x', t)$}
        \DisplayProof
    \end{subfigure}

    ~

    \begin{subfigure}[B]{0.45\textwidth}
        \centering
        \AxiomC{}
        \UnaryInfC{$union_\Xi(x, x) = \Xi, \bot$}
        \DisplayProof
    \end{subfigure}
    \begin{subfigure}[B]{0.45\textwidth}
        \centering
        \AxiomC{$x \neq y$}
        \AxiomC{$x \not\in dom(\Xi)$}
        \AxiomC{$y \not\in dom(\Xi)$}
        \TrinaryInfC{$union_\Xi(x, y) = [\Xi, (\{x, y\} \equiv x, \bot)], \bot$}
        \DisplayProof
    \end{subfigure}

    ~

    \begin{subfigure}[B]{0.9\textwidth}
        \centering
        \AxiomC{$[X \equiv x', t] \in \Xi$}
        \AxiomC{$x \in X$}
        \AxiomC{$y \not\in dom(\Xi)$}
        \TrinaryInfC{$union_\Xi(x, y) = [\Xi, (X, y \equiv x', t)], \bot$}
        \DisplayProof
    \end{subfigure}

    ~

    \begin{subfigure}[B]{0.9\textwidth}
        \centering
        \AxiomC{$x \not\in dom(\Xi)$}
        \AxiomC{$[Y \equiv y', t] \in \Xi$}
        \AxiomC{$y \in Y$}
        \TrinaryInfC{$union_\Xi(x, y) = union_\Xi(y, x)$}
        \DisplayProof
    \end{subfigure}

    ~

    \begin{subfigure}[B]{0.9\textwidth}
        \centering
        \AxiomC{$[X \equiv x', t_1] \in \Xi$}
        \AxiomC{$x \in X$}
        \AxiomC{$y \in X$}
        \TrinaryInfC{$union_\Xi(x, y) = \Xi, \bot$}
        \DisplayProof
    \end{subfigure}

    ~

    \begin{subfigure}[B]{0.9\textwidth}
        \centering
        \AxiomC{$[X \equiv x', t_1] \in \Xi$}
        \AxiomC{$x \in X$}
        \AxiomC{$[Y \equiv y', t_2] \in \Xi$}
        \AxiomC{$y \in Y$}
        \AxiomC{$x' \neq y'$}
        \QuinaryInfC{$union_\Xi(x, y) = [\Xi, (X \cup Y \equiv x', t_1)], (t_1, t_2)$}
        \DisplayProof
    \end{subfigure}
\end{figure}

Given definition of $union$ is left-biased (e.~g., w.~r.~t. representing variable and right hand side) but this does not affect the main properties and may be left implementation-dependent.

\subsection{The Unification Algorithm}

We define the unification algorithm $\emph{unify}_\Xi : T \times T \rightarrow \Xi_?$ for trees via the following inference system using the operations described before:

\begin{figure}[H]
    \centering
    \begin{subfigure}[B]{0.29\textwidth}
        \centering
        \AxiomC{$union_\Xi(x, y) = \Xi', \bot$}
        \RightLabel{(U1)}
        \UnaryInfC{$\emph{unify}_\Xi(x, y) = \Xi'$}
        \DisplayProof
    \end{subfigure}
    \begin{subfigure}[B]{0.36\textwidth}
        \centering
        \AxiomC{$union_\Xi(x, y) = \Xi', (t_1, t_2)$}
        \RightLabel{(U2)}
        \UnaryInfC{$\emph{unify}_\Xi(x, y) = \emph{unify}_{\Xi'}(t_1, t_2)$}
        \DisplayProof
    \end{subfigure}
    \begin{subfigure}[B]{0.33\textwidth}
        \centering
        \AxiomC{$\emph{find}_\Xi(x) = x', t'$ \quad $t' \neq \bot$}
        \UnaryInfC{$\emph{unify}_\Xi(x, t) = \emph{unify}_\Xi(t', t)$}
        \DisplayProof
    \end{subfigure}

    ~

    \begin{subfigure}[B]{0.95\textwidth}
        \centering
        \AxiomC{$\emph{find}_\Xi(x) = x', \bot$}
        \AxiomC{$y_1, ..., y_n$ --- fresh}
        \noLine
        \BinaryInfC{$\Xi_0 = bind_\Xi(x', f^n(y_1, ..., y_n))$ \quad $\Xi_1 = \emph{unify}_{\Xi_0}(y_1, t_1)$ \quad $...$ \quad $\Xi_n = \emph{unify}_{\Xi_{n - 1}}(y_n, t_n)$}
        \UnaryInfC{$\emph{unify}_\Xi(x, f^n(t_1, ..., t_n)) = \Xi_n$}
        \DisplayProof
    \end{subfigure}

    ~

    \begin{subfigure}[B]{0.53\textwidth}
        \centering
        \AxiomC{$\Xi_1 = \emph{unify}_{\Xi_0}(t_1, t_1')$ \quad $...$ \quad $\Xi_n = \emph{unify}_{\Xi_{n - 1}}(t_n, t_n')$}
        \UnaryInfC{$\emph{unify}_{\Xi_0}(f^n(t_1, ..., t_n), f^n(t_1', ..., t_n')) = \Xi_n$}
        \DisplayProof
    \end{subfigure}
    \begin{subfigure}[B]{0.45\textwidth}
        \centering
        \AxiomC{$f^n \neq g^m$}
        \UnaryInfC{$\emph{unify}_{\Xi_{~}}(f^n(t_1, ..., t_n), g^m(t_1', ..., t_m')) = \bot$}
        \DisplayProof
    \end{subfigure}
\end{figure}

For convenience we omit the handling of intermediate failures which terminate the algorithm as usual. The key differences from the classical unification algorithm (besides the lack of ``occurs check'') are:

\begin{itemize}
    \item an eager unification of variables by $union$ operation --- we first unify variables in an equation system and after that try to unify corresponding bounded trees;
    \item the conversion of complex trees to head form --- we decompose complex trees into a bunch of intermediate fresh variables and corresponding head trees.
\end{itemize}

These differences works in favor of the termination of the whole algorithm. To prove this, we first proof the termination for the case when both sides are variables. First of all, we define two auxiliary terms:

\begin{itemize}
    \item Scope --- the set of variables that at least includes all variables of the given equation system (both from left and right hand sides) and all variables in the unifying trees.
    \item Roots --- the subset of scope such that for every variable $x$ the condition $\emph{find}_\Xi(x) = x, t$ holds, i.~e. \emph{find} returns the same variable in the first component ($t$ may or may not be $\bot$).
\end{itemize}

\begin{coqlemma}
    The unification algorithm terminates for two variables.
\end{coqlemma}
\begin{proof}
    Induction on the number of roots.
    \begin{itemize}
        \item Base: no roots. Immediately implies an empty equation system, so the only possible case is (U1) that terminates immediately. Otherwise, let's consider an equation $X \equiv x, t \in \Xi$. The variable $x$ must be included in $X$ by definition of equation. Therefore it must be included in the domain of the equation system and in the scope. And it must be a root since $\emph{find}_\Xi(x) = x, t$ holds~--- a contradiction.
        \item Step: $n + 1$ roots. If we use (U1), we obviously terminate. If we use (U2), we may notice that given variables $x, y$ are different and must lay in different equivalence classes of the given system (i.~e. there must be two different roots $x', y'$ corresponding to them respectively). After $union$ a one of them must stop being a root and we don't add any new root so we may state that there are exactly $n$ roots in $\Xi'$ and the induction hypothesis applies.
    \end{itemize}
\end{proof}

\begin{coqlemma}
    The unification algorithm terminates.
\end{coqlemma}
\begin{proof}
    Well-founded induction on the maximum of the unifying tree heights. The base case is mostly covered by the previous lemma, the others just rely on the fact that the height of any right hand side is not greater than one.
\end{proof}

In order to prove the unifier property we prove a more general property that any unification produces a composition of variable substitutions to the set of infinite trees (using the notion of $image$ given before, we denote the interpretation of equation system $\Xi$ as substitution using $\llbracket\Xi\rrbracket_{image}$ notation) and it is a unifier of given trees.

\begin{coqlemma}
    Let $t_1, t_2$ be a pair of trees and $\Xi$ be an equation system. If the unification algorithm produces an equation system $\Xi'$, the following holds:
    \begin{itemize}
        \item there is a substitution $\sigma$ such that $\llbracket\Xi'\rrbracket_{image} = \sigma \circ \llbracket\Xi\rrbracket_{image}$;
        \item $t_1 \llbracket\Xi'\rrbracket_{image} = t_2 \llbracket\Xi'\rrbracket_{image}$ (unifier property).
    \end{itemize}
\end{coqlemma}
\begin{proof}
    Induction on the inference tree. First of all, note that the majority of basic operations trivially compose an existing equation system with some new unifying binding. The only exception is the second rule, when we rewrite an existing binding $x \mapsto image_\Xi(x) \llbracket\Xi\rrbracket_{image}$ with \mbox{$x \mapsto image_\Xi(y) \llbracket\Xi\rrbracket_{image}$}.

    In this case, we must note that after a successful unification of $t_1$ and $t_2$ we may state that in the resulting equation system (named $\Xi'$) they are unified, so $image_{\Xi'}(x) \llbracket\Xi'\rrbracket_{image} = image_{\Xi'}(y) \llbracket\Xi'\rrbracket_{image}$ holds. Moreover, preemptive union of variables doesn't change the substitution interpretation.
\end{proof}

While we expect the unifier delivered by our algorithm to be the most general we do not have a formal proof yet. Moreover, strictly speaking as we introduce fresh variables in order to convert initial unification trees into head tree form and these variables can later be bound in the answer the unifier \emph{may not be} the most general. We think this subtlety can be worked around.

For example, unifying $x = f(g(y))$ leads to adding intermediate fresh variable $z$ to construct system $\{ \{ x \} \equiv f(z); \{ z \} \equiv g(y) \}$ that will act like substitution $\{ x \mapsto f(g(y)) ; z \mapsto g(y) \}$. This system does not possess the MGU property since the ``real'' MGU is $\{ x \mapsto f(g(y)) \}$ --- note the absence of mapping for $z$. So, to have the MGU property we must not take inro account fresh variables.

\subsubsection{Example 1}

For non-recursive equations our algorithm works similarly to the conventional one except for adding fresh variables. Let's consider a simple example: $f(x, g()) = f(u, v)$.
\begin{enumerate}
    \item Initially we have an empty equation system: $\Xi_0 = \varnothing$.
    \item Since the functional symbols are the same, we decompose given equation into two: $x = u$ and $g() = v$.
    \item For the equation $x = u$ we compute $union_{\Xi_0}(x, u)$. It gives us a new system $\Xi_1 = \{ \{ x, u \} \equiv x, \bot \}$ and doesn't return a pair of head trees so we just go on with $\Xi_1$.
    \item For the second one we first compute $\emph{find}_{\Xi_1}(v) = v, \bot$.
    \item Since $v$ isn't bound to any head tree we bind $v$ with $g()$ so $\Xi_2 = \{ \{ x, u \} \equiv x, \bot ; \{ v \} \equiv v, g() \}$
    \item The resulting equation system is:
    \begin{equation*}
        \Xi_2 = \left\{\begin{aligned}
            x, u &\equiv x, \bot \\
            v &\equiv v, g() \\
        \end{aligned}\right.
    \end{equation*}
    \item It corresponds to the following substitution:
    \begin{equation*}
        \llbracket\Xi_2\rrbracket_{image} = \left\{\begin{aligned}
            x &\mapsto x \\
            u &\mapsto x \\
            v &\mapsto g() \\
        \end{aligned}\right.
    \end{equation*}
\end{enumerate}

As we can see, in simple cases our algorithm works similarly to the conventional one. It's easy to see that in the case of different functional symbols in the both sides of given equation it terminates with a failure like in the conventional algorithm.

\subsubsection{Example 2}

For recursive equations we produce recursive system of equations. Let's consider a case $x = f(g(x))$:
\begin{enumerate}
    \item Initially we have an empty system: $\Xi_0 = \varnothing$.
    \item First, compute $\emph{find}_{\Xi_0}(x) = x, \bot$.
    \item As in the previous example we need to bind $x$ to a head tree. In order to do it, we introduce a fresh variable $x_1$ and do $bind_{\Xi_0}(x, f(x_1)) = \{ \{x\} \equiv x, f(x_1) \} = \Xi_1$.
    \item Next, we need to unify $x_1$ with $g(x)$.
    \item Like before, we introduce a fresh variable $x_2$ and bind:
    $bind_{\Xi_1}(x_1, g(x_2)) = \{ \{x\} \equiv x, f(x_1) ; \{x_1\} \equiv x_1, g(x_2) \} = \Xi_2$.
    \item Now, we are going to unify $x_2$ with $x$.
    \item We compute $union_{\Xi_2}(x_2, x) = \Xi_3, \bot$, where $\Xi_3 = \{ \{x\underline{, x_2}\} \equiv x, f(x_1) ; \{x_1\} \equiv x_1, g(x_2) \}$.
    \item Since $union$ hasn't return a pair of head trees, we stop here.
    \item The result is:
    \begin{center}
        \begin{minipage}[c]{0.3\textwidth}
            \begin{equation*}
                \Xi_3 = \left\{\begin{aligned}
                    x, x_2 &\equiv x, f(x_1) \\
                    x_1 &\equiv x_1, g(x_2) \\
                \end{aligned}\right.
            \end{equation*}
        \end{minipage}
        \begin{minipage}[c]{0.55\textwidth}
            \begin{equation*}
                \llbracket\Xi_3\rrbracket_{image} = \left\{\begin{aligned}
                    x &\mapsto f(g(f(g(...)))) \\
                    x_2 &\mapsto f(g(f(g(...)))) \\
                    x_1 &\mapsto g(f(g(f(...)))) \\
                \end{aligned}\right.
            \end{equation*}
        \end{minipage}
        \begin{equation*}
            \llbracket\Xi_3\rrbracket_{\mu\text{-}image} = \left\{\begin{aligned}
                x &\mapsto \mu x. ~ f(g(x)) \\
                x_2 &\mapsto \mu x. ~ f(g(x)) \\
                x_1 &\mapsto \mu x_1. ~ g(f(x_1)) \\
            \end{aligned}\right.
        \end{equation*}
    \end{center}
\end{enumerate}

Here we can see that the result system interpretation corresponds to the result of conventional unification with no ``occurs check'' modulo
fresh variables introduction.

\subsubsection{Example 3}

Finally, let's consider an example when the conventional algorithm without ``occurs check'' loops infinitely: $f(x, g(x)) = f(g(g(x)), x)$.
\begin{enumerate}
    \item Initially we have an empty system: $\Xi_0 = \varnothing$.
    \item Decompose given equation into the two: $x = g(g(x))$ and $g(x) = x$.
    \item For the first one the solution is similar to the previous example:
    \begin{equation*}
        \Xi_1 = \left\{\begin{aligned}
            x, x_2 &\equiv x, g(x_1) \\
            x_1 &\equiv x_1, g(x_2) \\
        \end{aligned}\right.
    \end{equation*}
    \item For the second, we first compute $\emph{find}_{\Xi_1}(x) = x, g(x_1)$.
    \item Since $x$ is already bound to a head tree $g(x_1)$ we unify $g(x)$ with it.
    \item Simplify the equation to $x = x_1$.
    \item Compute $union_{\Xi_1}(x, x_1) = \Xi_2, (g(x_1), g(x_2))$, where $\Xi_2 = \{ \{ x, x_2, x_1 \equiv x, g(x_1) \} \}$.
    \item Since $union$ has returned a pair $g(x_1), g(x_2)$, we need to unify these head trees.
    \item Simplify to $x_1 = x_2$.
    \item Compute $union_{\Xi_2}(x_1, x_2) = \Xi_2, \bot$.
    \item Since $x_1$ and $x_2$ are already presented in the same equation, union does nothing and we stop.
    \item The result is:
    \begin{equation*}
        \Xi_2 = \left\{\begin{aligned}
            x, x_2, x_1 &\equiv x, g(x_1) \\
        \end{aligned}\right.
    \end{equation*}
    \begin{center}
        \begin{minipage}[c]{0.35\textwidth}
            \begin{equation*}
                \llbracket\Xi_2\rrbracket_{image} = \left\{\begin{aligned}
                    x &\mapsto g(g(...)) \\
                    x_2 &\mapsto g(g(...)) \\
                    x_1 &\mapsto g(g(...)) \\
                \end{aligned}\right.
            \end{equation*}
        \end{minipage}
        \begin{minipage}[c]{0.55\textwidth}
            \begin{equation*}
                \llbracket\Xi_2\rrbracket_{\mu\text{-}image} = \left\{\begin{aligned}
                    x &\mapsto \mu x. ~ g(x) \\
                    x_2 &\mapsto \mu x. ~ g(x) \\
                    x_1 &\mapsto \mu x. ~ g(x) \\
                \end{aligned}\right.
            \end{equation*}
        \end{minipage}
    \end{center}
    \item In contrast to our algorithm the conventional one will loop infinitely on the series of equations like
    \[g(x) = x \Rightarrow g(x) = g(g(x)) \Rightarrow x = g(x) \Rightarrow g(g(x)) = g(x) \Rightarrow g(x) = x \Rightarrow ...\]
\end{enumerate}

\subsection{Answer minimization}

As we mentioned before, we are studying answer minimization problem. For example, we want to get an answer $\mu x. ~ x \rightarrow y$ instead of $(\mu x. ~ x \rightarrow y) \rightarrow y$ since they represent same rational trees but the first one is simpler. This problem was considered in~\cite{courcelle1974algorithms} where an algorithm that produces a minimal representation for any equation system by union subtrees with equivalent images was presented. When we have an equation system with non-head trees in its right hand side we must enumerate all subtrees to perform minimization. Obviously, it's simpler to enumerate only variables and restricting the right hand sides to be head trees makes it possible.

For example, let us consider the following equation system:
\begin{equation*}
    \left\{
    \begin{split}
        \{ x \} &\equiv f(x, g(x, y)) \\
        \{ y \} &\equiv g(f(x, y), y) \\
    \end{split}
    \right.
\end{equation*}
The $\mu$-image of $x$ is $\mu x. ~ f(x, g(x, \mu y. ~ g(f(x, y), y)))$.

To enumerate all subtrees we need to walk the right hand sides: $x$, $y$, $g(x, y)$, $f(x, g(x, y))$, $f(x, y)$, $g(f(x, y), y)$. So, let's introduce fresh variables to make right hand sides head trees:
\begin{equation*}
    \left\{
    \begin{split}
        \{ x \} &\equiv f(x, x') \\
        \{ x' \} &\equiv g(x, y) \\
        \{ y \} &\equiv g(y', y) \\
        \{ y' \} &\equiv f(x, y) \\
    \end{split}
    \right.
\end{equation*}

Now we may enumerate only left hand sides $x$, $x'$, $y$, $y'$ since all right hand sides are equivalent to some of them. Next, we may apply an algorithm from Lemma 10 of $\cite{courcelle1974algorithms}$ to minimize it:
\begin{equation*}
    \left\{
    \begin{split}
        \{ x, y' \} &\equiv f(x, x') \\
        \{ y, x' \} &\equiv g(y', y) \\
    \end{split}
    \right.
\end{equation*}
Now, the $\mu$-image of $x$ is just $\mu x. ~ f(x, \mu y. ~ g(x, y))$.

\subsection{Implementation Notes}

\subsubsection{Equation System}

We use the ideas of union-find data structure~\cite{arden1961algorithm} to represent equation systems in memory. Generally, an equation of the form $V \equiv v, t$ is represented by a mapping from each of $x \in V, x \neq v$ to some variable $y \in V$ such that no circular mappings like

\[x_1 \mapsto x_2 \mapsto ... \mapsto x_n \mapsto x_1 \mapsto ... ,\]

exist.

We call these variables ``linked''. At this time, the representing variable $v$ may be mapped to the head tree $t$ or may not be mapped at all.

The most straightforward implementation of the algorithm relies on the reusing of data structure of classical variable substitution of type \mintinline{OCaml}{Tree.t Tree.VarMap.t} that is a mapping from variables to trees. In this implementation we restrict the right hand side of a mapping to be either a variable or a head tree.

The possible improvement would be adding a path compression as in conventional union-find. To achieve this, we can make the data structure mutable: \mintinline{OCaml}{Tree.t Tree.VarMap.t ref}.

The next improvement is adding a rank heuristic by replacing the data type for system representation with \break \mintinline{OCaml}{node Tree.VarMap.t ref}, where \texttt{node} is defined as

\begin{minted}{OCaml}
type root = {
  depth: int;
  tree: tree.t option;
}

type node =
| RootNode of root
| LinkNode of tree.Var.t
\end{minted}

We will recall these three flavors of the core implementation as ``classical'', ``path-compression'' and ``union-find'' respectively. Actually we may consider these flavors w.~r.~t. the classical unification algorithm, too, and we have implemented them for evaluation purposes.

\subsubsection{Attributed Variables}

A.k.a. ``\texttt{set-var-val!}'', ``breaking assignment'', etc.

This optimization is based on tracking the scopes of fresh variables and is used in the original \textsc{OCanren}. We use it, too, in our implementation. Although it's trivially applies for ``classical'' flavor, the others are tricky.

For path-compression flavor we aren't allowed to do path compression for attributed variables since compression is based on the given equation system. It could lead to a long paths of bindings that we cannot compress and nullify the advantages of path compression. On the other hand, we may write such compressed bindings in a mapping but this will nullify the advantages of attributed variables in variables lookup.

For union-find flavor we cannot write ranks as parts of variable attribution since they are local for the given equation system, so we just assume them to be zero. It nullifies the advantages of union-find data structure, too.

\subsubsection{Variable Comparing}

While experimenting with union-find flavor we have noticed that variable numbers in produced answers became smaller.

It could be explained by the linearity of variable number allocation or simply ``greater numbers are allocated later''. Intuitively, a variable's rank correlates with it's life time that is correlated with it's number.

This leads us to the idea of variable numbers comparing during unification. Instead of left-biased binding we can always bind the greater variable to the less one. For example, if we do unification of $\_.1$ and $\_.2$, we populate the mapping with $\_.2 \mapsto \_.1$ instead of $\_.1 \mapsto \_.2$. This technique is applicable for both classical and our unification algorithms.

\subsubsection{Disequality Constraints}

As it was mentioned above, our algorithm doesn't deliver the MGU property in general. This leads to a problem with disequality constraint solving since we cannot state that disunification solves disequality.

To address this, we collect a unification prefix without introduced fresh variables, as in classical unification algorithm. For example, during the unification of $\_.1$ with $f(g(\_.2))$ we extend the equation system with bindings $\_.1 \mapsto f(\_.3)$ and $\_.3 \mapsto g(\_.2)$ but return the prefix $\_.1 \mapsto f(g(\_.2))$. It fixes the MGU property \emph{ad hoc} and disunification solving.

Another problem is collecting prefix under union-find flavor with attributed variable optimization. Since we may use $union$ operation of union-find only for non-local pairs of variables, we must decouple $union$ of union-find from $union$ of our equation system interface. It leads to a problem that we actually don't know what variable binds to another of two unified and could cause inconsistent prefix production. So the correct implementation is actually very tricky.

\subsubsection{Occurs Check}

An interesting question is how can we implement the unification with occurs check using our algorithm. Since we aren't required to do occurs check on every step of unification we may consider different ways to delay occurs check. We have implemented several occurs check strategies:

\begin{itemize}
    \item Trivial --- similar to the classical algorithm the occurs check is performed before extending an equation system. It maintains the equation system in a non-recursive state (forbidding infinite images), so it actually could be just a plain old substitution.
    \item Simple --- occurs check is performed after every unification. We check variables from a collected unification prefix using a simple DFS-algorithm to find cycles.
    \item Full --- occurs check is performed before producing answers. We check the whole equation system right before providing an answer to an end-user. Since a number of variables have their bindings as attributes with attributed variable optimization, we additionally record attributed variables to check them, too.
    \item Multiplicative --- similar to the ``full'' strategy but additionally runs full occurs check after every $m$ unifications. The value $m$ is doubled after every occurs check (like in a multiplicative vector allocation strategy). Intermediate occurs checks don't check attributed variables separately since the cycles on them only are rare.
    \item SQRT --- a symbiosis of ``multiplicative'' and ``simple'' strategies. We don't do occurs check after every unification but after every $m$ unifications and at the end we do the full occurs check. Unlike ``multiplicative'' strategy, the intermediate checks aren't full but for recorded set of bound variables since previous occurs check. The variable $m$ is adjusted after every occurs check, too, but as a square root of the current size of the equation system. The idea is based on ``SQRT-decomposition''\footnote{The original inventor is unknown, for example this technique could be studied from \url{https://cp-algorithms.com/data_structures/sqrt_decomposition.html}} technique for algorithm construction.
\end{itemize}

Note that while we won't hang while unification using our algorithm, unexpected hangs are possible from relational program. For example, consider the query $\exists x. x = f(x) \land \texttt{loop}^o()$ where ``$\texttt{loop}^o$'' is defined like $\top \lor \texttt{loop}^o()$ i.~e. doesn't unify anything and infinitely produces answers. Using ``trivial'' strategy or non-rational unification algorithm we will stop evaluation right after first conjunct but with occurs check delaed we will hang infinitely without any answer. It could be worked around by using a different method to count ``steps'' of evaluation (e.~g. counting relation calls along with unifications) but it generally isn't an issue.

\section{Evaluation}

% subcaption increases counter even for figures without captions
\setcounter{figure}{0}

In this section we present the evaluation of our rational unification implementation and its flavors using well-known classical samples and some samples that require rational tree support. Additionally, we discuss the notions of quines in the presence of recursively infinite programs.

First of all, the implementation has passed all \textsc{OCanren} regression tests. The answers mostly stay same (w.~r.~t. variable renaming) except for obvious cases when conventional occurs check gets in the way. For instance, STLC type inference now successfully infers a recursive type $(\mu x. ~ x \rightarrow y) \rightarrow y$ for the term with self application $\lambda x. ~ x ~ x$.

\begin{figure}
    \centering
    \includegraphics[height=0.35\paperheight]{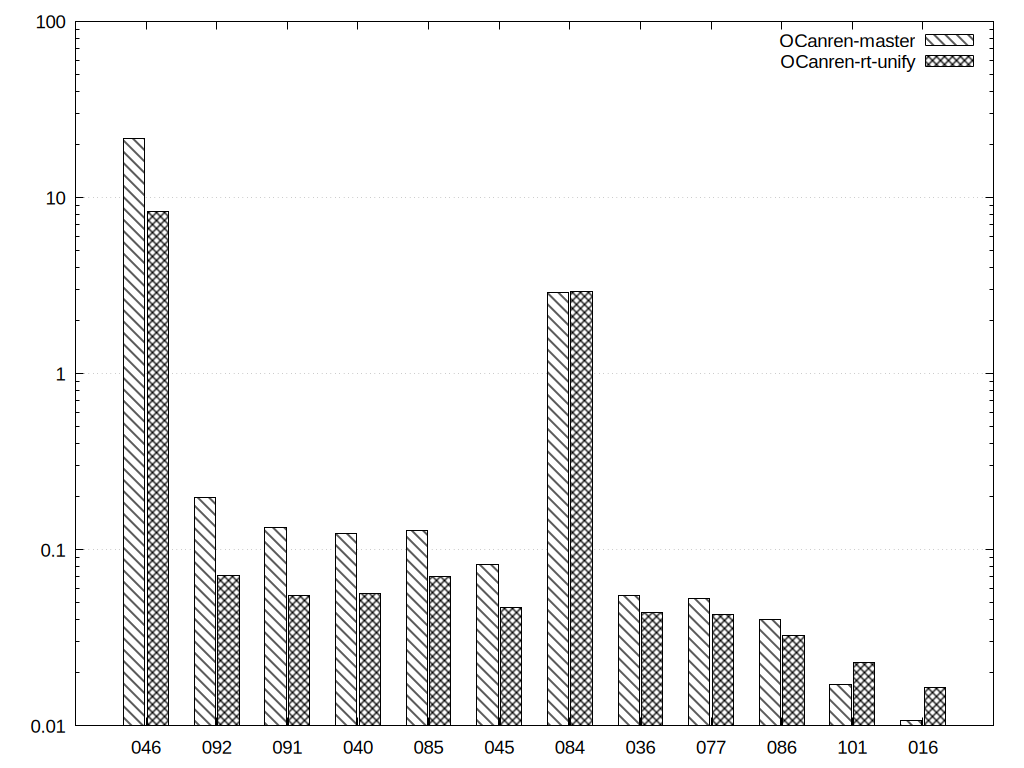}
    \caption{Solver for equirecursive type inference with conventional and rational tree unification}
    \label{fig:lama}
\end{figure}

We have modified the type constraint solver from~\cite{domoratskiy2024relational} to use rational tree unification in place of ``occurs hooks''. It has simplified the whole implementation and increased it's speed in the worst case from 21 sec. to 8 sec. At the same time, other tests were not significantly affected (indicating a differences about a tenths of second). The inferred types stay the same as for the original implementation, too, for all successful runs. The results of benchmarking for some notable tests are presented in figure~\ref{fig:lama}. We use a logarithmic scale here because of the great difference between the first test and the rest.

The main drawback of our \textsc{OCanren} implementation is its inability of simple writing a structurally recursive relation definitions over rational trees since they always hang by infinite recursion. This discriminates our implementation from a complete coinductive logic programming~\cite{simon2006coinductive} that provides coinductive relation definitions support. Because of this we have excluded two tests that rely on structurally recursive-defined relations.

Another issue with the original implementation is that it doesn't minimize the producing answers and our doesn't do it too. We have studied that there is an algorithm~\cite{courcelle1974algorithms} that provides an ability to minimize answers but haven't implemented it yet for now. The main point is that this algorithm requires equation systems to be in so-called ``canonical form''. This form requires that any equation must have a head tree in its right hand side and  our unification algorithm maintains it, too. So it must be pretty simple to implement the minimization as a final step before producing answers.

Additionally we have evaluated our implementation on the more conventional problems:

\begin{itemize}
    \item \texttt{001} --- $exp^o(3, 5, x)$;
    \item \texttt{002} --- $exp^o(3, x, 243)$;
    \item \texttt{005} --- two thrines;
    \item \texttt{006} --- 30 twines;
    \item \texttt{007} --- 200 quines;
    \item \texttt{011} --- 200 quines without disequality.
\end{itemize}

\begin{figure}
    \centering
    \includegraphics[height=0.35\paperheight]{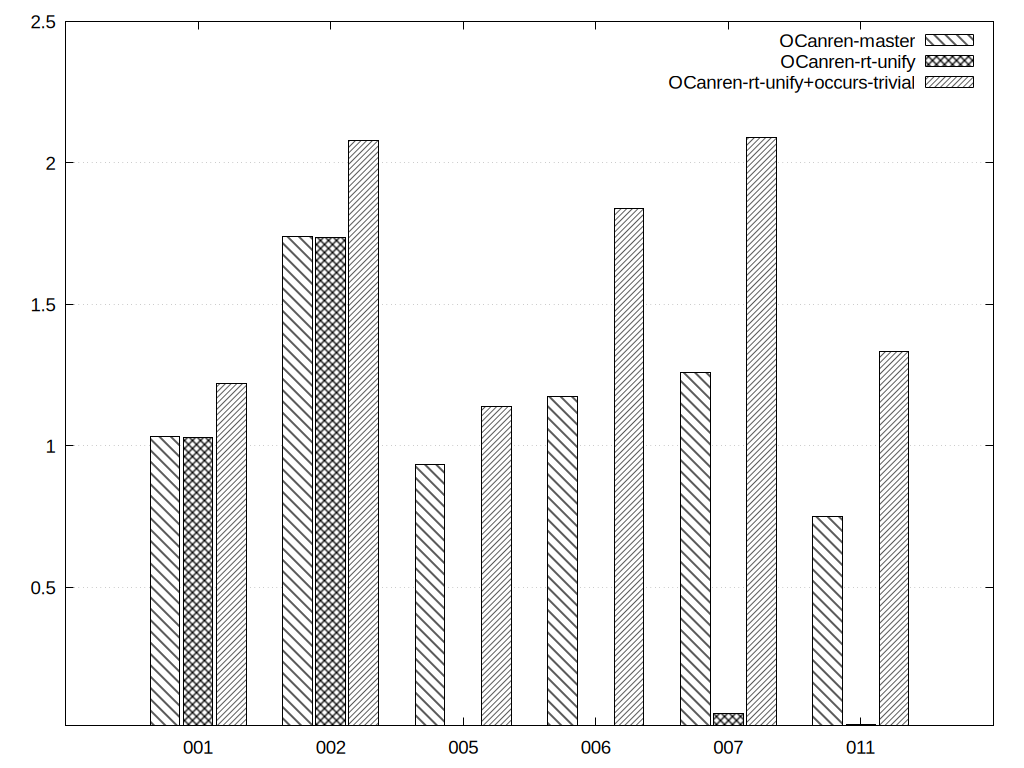}
    \caption{Conventional vs. rational vs. rational with occurs check}
    \label{fig:master-rt-bench}
\end{figure}

For the first two tests \textsc{OCanren} with rational tree unification gives the same answers as the original since they don't rely on occurs check. For the rest, our implementation primarily finds programs with recursive bodies as expected. The simplest quine is $\mu x. ~ (\texttt{quote} ~ x)$. When we add occurs check  the answers become similar to the conventional ones. The benchmarking results are presented in figure~\ref{fig:master-rt-bench} for the conventional \textsc{OCanren}, our implementation without occurs check and our implementation with ``trivial'' occurs check. Our implementation was used in the simplest variant with ``classical'' equation system representation and without variable comparing optimization.

We may notice the higher evaluation time for our implementation with occurs check than for the conventional one. It must be because of the overhead from decomposing complex trees in a head trees and adding new variables. Obviously, the variant without occurs check is finished faster for the quine tests since it didn't forbid cyclic answers.

The rest of measurements will be presented only for the variant with occurs check to compare them with the conventional implementation.

\subsection{Equation System Representation}

\begin{figure}
    \centering
    \includegraphics[height=0.35\paperheight]{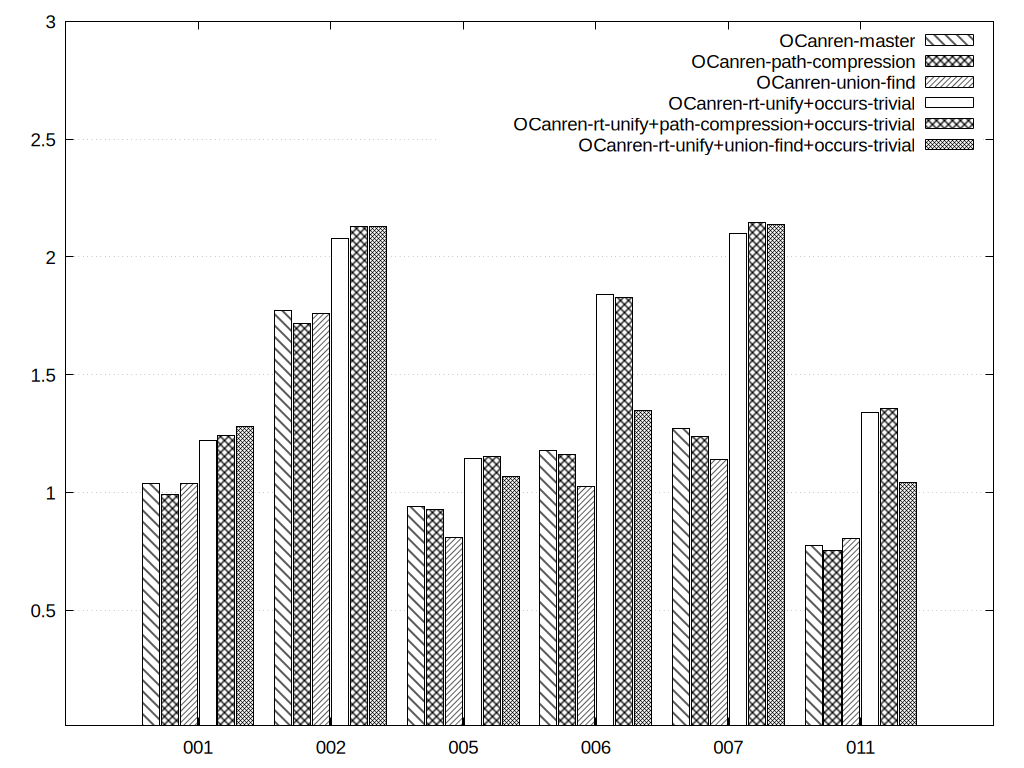}
    \caption{Conventional and rational tree unification with different substitution/equation system representations}
    \label{fig:eqsys-flavs}
\end{figure}

As it was mentioned above, we have implemented three flavors of equation system representation: classical, path-compression and union-find. Additionally, we have implemented the conventional \textsc{OCanren} with substitution representation in the similar flavors. The benchmarking results are presented in  figure~\ref{fig:eqsys-flavs}.

Although union-find representation sometimes helps us, it could sometimes slow down the process, too, because of the greater overhead of maintaining auxiliary data structures. Since potential performance advantage isn't so great, we have decided not to use it. A similar situation is with path compression representation. Its overhead isn't so great but the profit isn't significant, too. The implementation difficulties mentioned before only make the situation worse.

\subsection{Attributed Variable Optimization}

\begin{figure}
    \centering
    \includegraphics[height=0.35\paperheight]{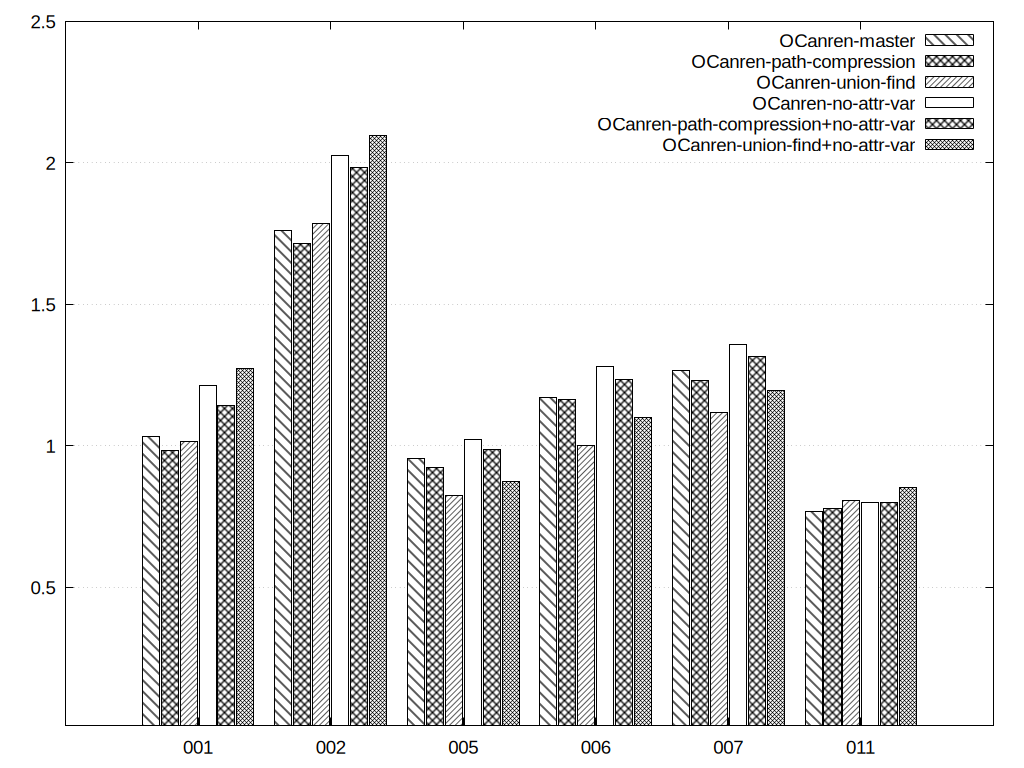}
    \caption{Conventional unification with and without attributed variable optimization}
    \label{fig:no-attr-var-master}

    \vskip1em

    \includegraphics[height=0.35\paperheight]{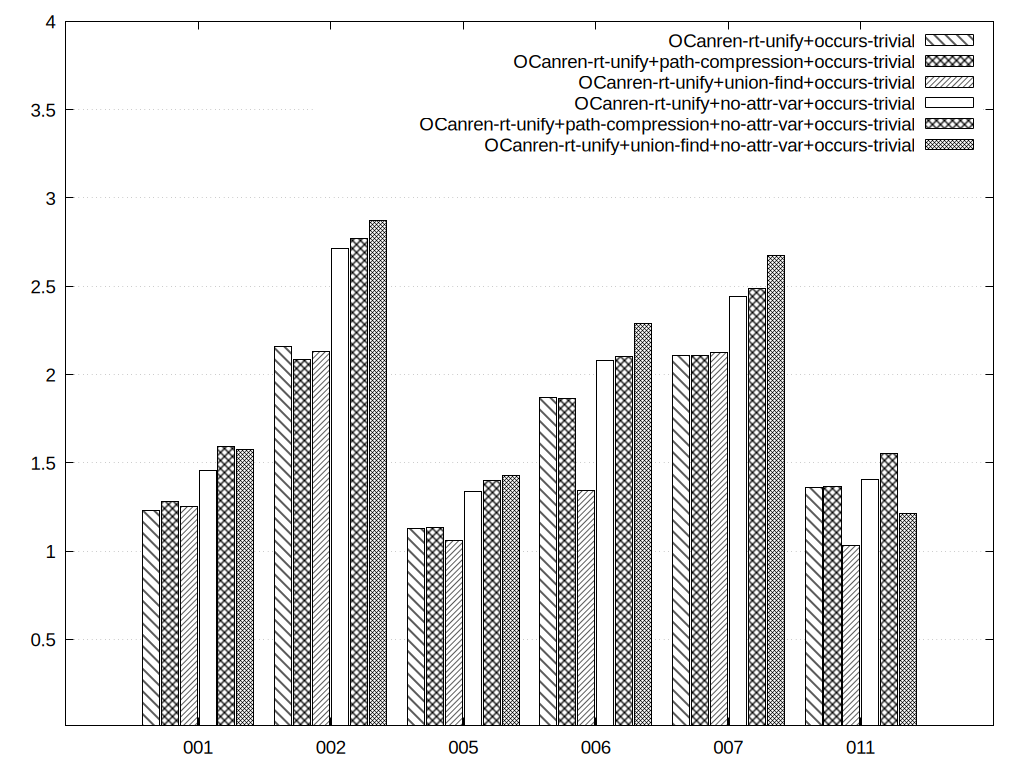}
    \caption{Rational unification with and without attributed variable optimization}
    \label{fig:no-attr-var-rt}
\end{figure}

We have additionally implemented variants of \textsc{OCanren} without attributed variable optimization mentioned before for all of described flavors of equation system and substitution (for our algorithm and for the conventional, respectively). The measurements are presented in figure~\ref{fig:no-attr-var-master} and figure~\ref{fig:no-attr-var-rt}.

We may see that attributed variable optimization significantly affects the performance in the worst case for both conventional unification and rational tree unification. Taking in account its incompatibility with advanced flavors of substitution (or equation system) implementation discussed above it looks like the most reasonable way is to use the plain old mapping without any improvements.

\subsection{Variable Comparing Optimization}

\begin{figure}
    \centering
    \includegraphics[height=0.35\paperheight]{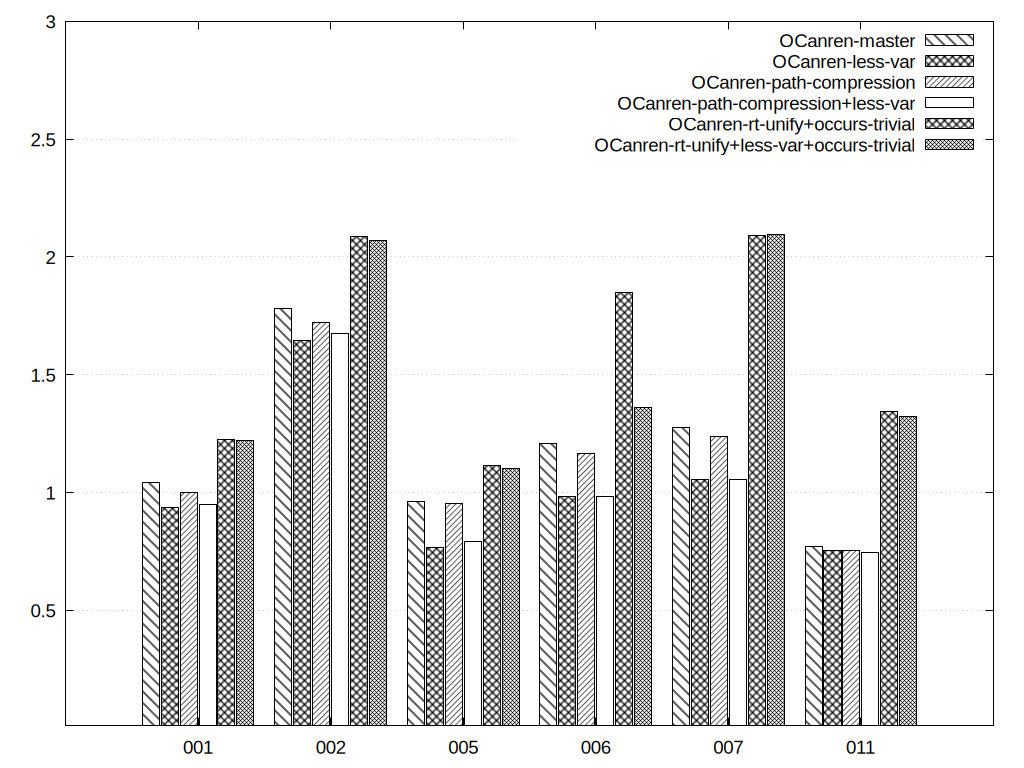}
    \caption{\textsc{OCanren} variants with and without variable comparing and path compression}
    \label{fig:less-var}
\end{figure}

We have also implemented a few variants of \textsc{OCanren} with variable comparing optimization discussed before. We have considered three base variants:  conventional, conventional with path compression to compare performance gain, and with our unification algorithm. The benchmarking results are presented in figure~\ref{fig:less-var}.

As we may see, variable comparing optimization never reduces the performance and frequently has a positive performance impact comparable with path compression. It's achieved because of about zero overhead of this optimization (one compare on variable-variable unification) and by the theoretical reasons explained before. It could be considered as a lightweight replacement for rank heuristic from union-find data structure but specialized for unification algorithms. As a nice bonus we have reduced variable numbers that simplify answer understanding.

\subsection{Occurs Check Strategy}

\begin{figure}
    \centering
    \begin{subfigure}{0.9\linewidth}
        \centering
        \includegraphics[width=\linewidth]{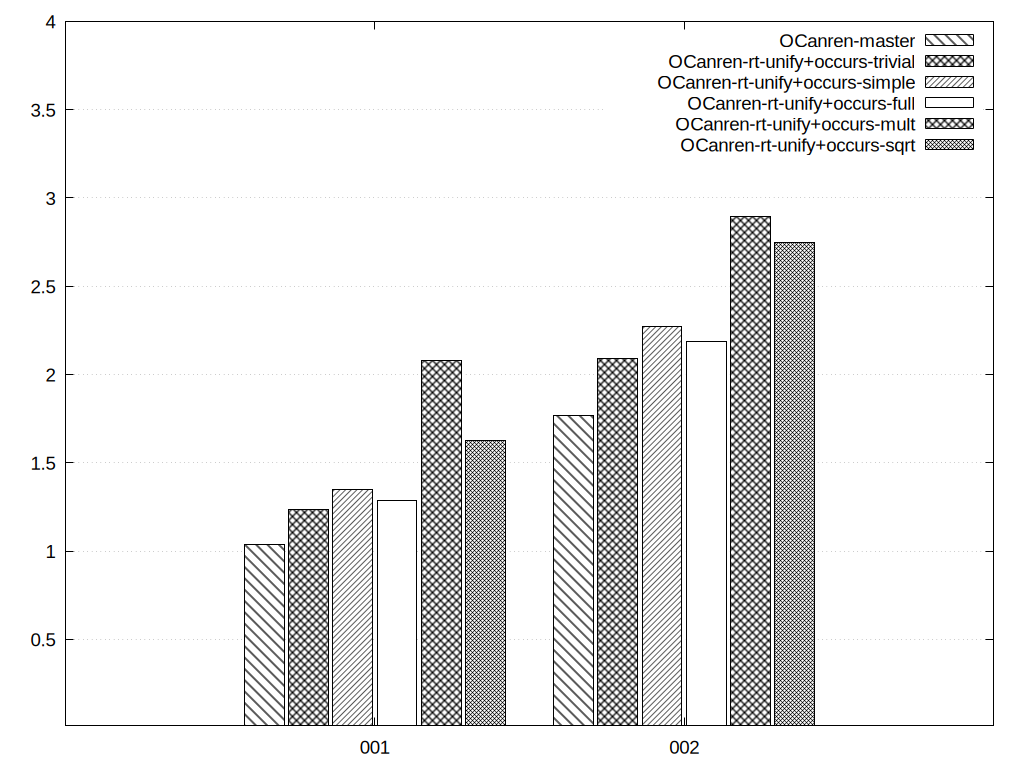}
        \caption{Simple tests}
    \end{subfigure}

    \vskip1em
    
    \begin{subfigure}{0.4\linewidth}
        \centering
        \includegraphics[width=\linewidth]{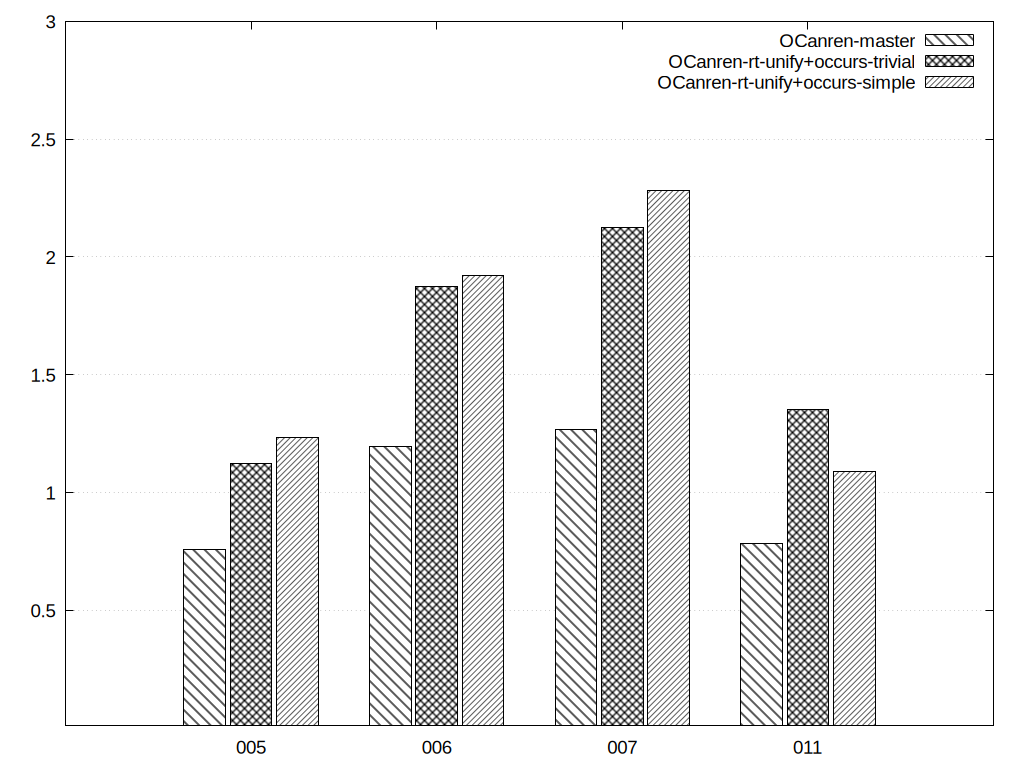}
        \caption{Simplest strategies on quine tests}
    \end{subfigure}
    \quad
    \begin{subfigure}{0.4\linewidth}
        \centering
        \includegraphics[width=\linewidth]{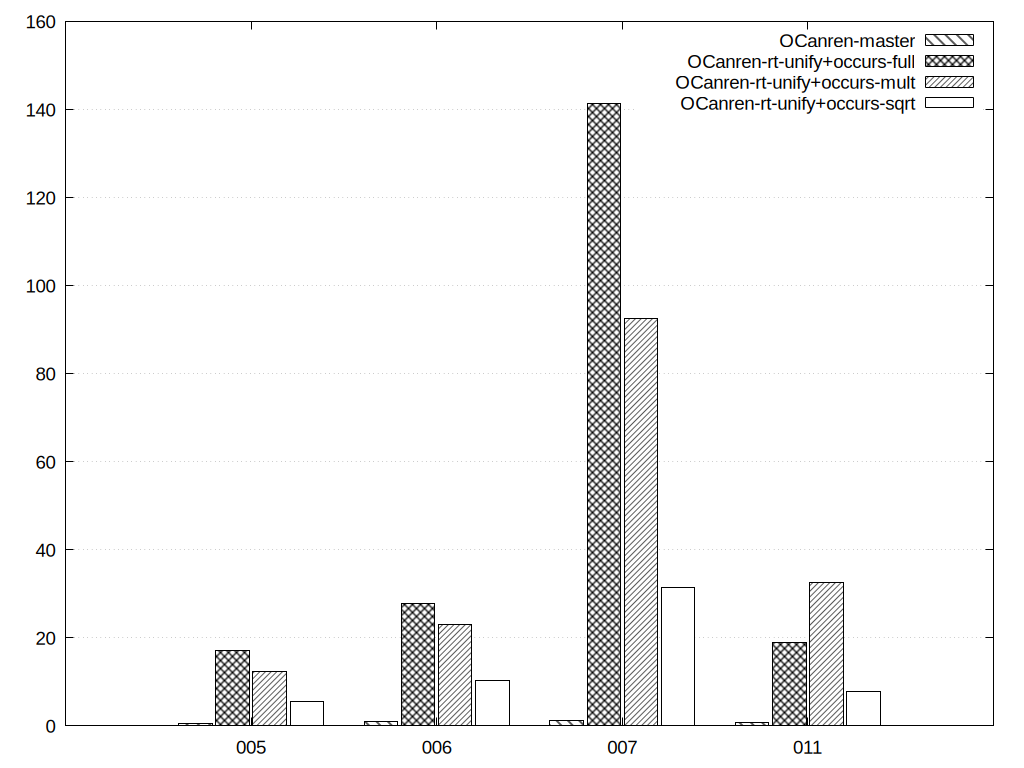}
        \caption{Other strategies on quine tests}
    \end{subfigure}
    
    \caption{Different occurs check strategies used with rational tree unification}
    \label{fig:occurs-check}
\end{figure}

Finally, we have implemented five different strategies to perform occurs check described above. We have followed an intuition that in cases of programs that don't rely on occurs check, we may increase the performance by delaying occurs check up to the answer yielding. And to make it tolerant to cases that rely on occurs check we have considered different ways of performing occurs check once in a several unifications. The final results are turned out to be interesting. We present them in figure~\ref{fig:occurs-check}.

We are splitting the results on the several plots because of a great difference in scale needed to observe them. On the first plot we show the results for  two simple tests that don't rely on occurs check. As expected, the ``full'' strategy works good since we don't need to eliminate many search branches. But it doesn't work better that the ``trivial'' strategy. So even for the simplest cases the simplest is the best.

On the last two plots we show results for quine-related tests. On the first one there are ``trivial'' and ``simple'' strategies, and others are on the second one. We may see that ``trivial'' strategy works better than ``simple''. It could be caused by some overhead of using DFS with remembering of visited variables for the ``simple'' strategy. And both of them work much better than the rest of strategies showed on the last plot.

As expected, the ``full'' strategy is the worst among all other strategies and ``multiplicative'' isn't much better. It is interesting that the last strategy  ``SQRT'' works much better than they, but much worse than the simplest strategy.

We think that any delaying occurs check strategy must work worse than non-delaying in \textsc{miniKanren} because of non-determinism. Any delaying strategy collects its work from several steps every time and must do it in a several branches multiple times instead of single time when we do it immediately, so we slow down the whole search. At the same time, delayed elimination of invalid search branches slows down the search because of interleaving search that consumes answers ``in parallel'' from all existing branches, so invalid branches interfere with valid ones .

\section{Related Work}

Unification is a well-studied area in computational logic, and the unification of rational trees is no exception. There is a number of works~\cite{martelli1984efficient,huet1976resolution,mukai1983unification} presenting unification algorithms for rational tree unification with the proofs of their termination and correctness. The majority of them are defined as iterative algorithms that rewrite an equation system until it becomes solved. They require complicated data structures to support equation system such as ``system of multiequation'' that are useful for proving algorithm properties but difficult to implement efficiently. We have replaced the data structure to represent equation system to formulate unification algorithm as recursively descending like classical unification algorithms and proved it's termination and unifier properties directly.

Some of the works~\cite{martelli1984efficient} additionally discuss rational tree unification with occurs check as a possible replacement for classical unification algorithms. One of the options is to support additional tags to detect cycles without explicit occurrence search. Also, there are works~\cite{pearce2007dynamic} that present dynamic algorithms to support acyclic graph structures. We aren't aware of any prior work on empirical study of different occurs check strategies such as delayed occurs check.

Representation of equation system (or substitution) as union-find~\cite{arden1961algorithm} data structure is described in many works, primarily~\cite{martelli1984efficient} for rational tree unification. The persistent variant of union-find based on persistent mapping is well known, but there is a work on another implementation~\cite{conchon2007persistent}. However, this result isn't applicable to \textsc{miniKanren} because of interleaving search.

Finite representation of rational trees using $\mu$-notation comes from type theory and is well-known~\cite{pierce2002types}. The problem of finite representation minimization was studied independently of $\lambda$-calculus by Courcelle~et.~al.~\cite{courcelle1974algorithms}. They presented an algorithm to find an equivalence between regular trees (that is different name for rational trees). This algorithm is based on enumeration of subtrees and could be simplified for equation systems in ``canonical'' form, that is essentially a form that is produced by the our unification algorithm. The simplification is based on the fact that the right hand side of every equation in system is head tree, i.~e. there are no subtrees except for variables.

\section{Conclusion and Future Work}

We have presented an unification algorithm for rational trees that is technically could be used as a replacement for the conventional unification algorithm to support recursive data structures in \textsc{miniKanren}. We have proven it's termination and unifier properties but MGU property doesn't hold in general because of fresh variable introduction. In practice, this issue could be worked around, for example, to implement disequality solving.

We have implemented our algorithm in \textsc{OCanren} and evaluated it for many examples. For a solver that requires rational tree support it helps us to gain performance improvement in comparison with an \emph{ad-hoc} workaround. Unfortunately, the overhead produced by the algorithm makes it unsuitable to be a drop-in replacement for the classical algorithm yet.

We have studied several implementations for substitution and equation system representations and demonstrated that path compression technique and union-find data structure don't increase the performance significantly but complicate the implementation of \textsc{miniKanren} and introduce additional overhead.

Additionally, we have presented a novel optimization heuristic named ``variable comparing'' for the classical unification algorithm based on study of rank heuristic of union-find data structure. We have evaluated it and showed that it could increase the performance in many cases without any drawbacks and compared it with path compression technique. It helped us to show that our heuristic is better than path compression since it increases the performance greater and the simultaneous usage of ``variable comparing'' and path compression doesn't increases the performance significantly. But our optimization is much simpler to implement.

Also, we have measured a performance gain given by a well-known attributed variable optimization and showed that it increases the performance significantly. So it cannot be omitted even with the usage of more advanced substitution implementations such as union-find data structure.

Finally, we have studied several occurs check application strategies including delayed ones and showed that any non-constant delaying of occurs check decreases the performance significantly, so any solver that uses occurs check must apply it on every unification as in the conventional algorithm.

We understand the irrelevance of the presented algorithm for practical usage and are going to work on other algorithms that could provide rational tree unification without such a great overhead. As was mentioned before, rational unification isn't applicable to real applications without coinductive relation definition support since it doesn't allow to operate with rational trees being in input position that causes infinite looping. So in the future we are going to work on coinductive relations support for \textsc{miniKanren}, too.

And the last but not least, we consider a possibility to use rational trees with Herbrand trees simultaneously to provide an ability to deal with data structures like ``stream of lists'' where ``list'' is a finite data structure and ``stream'' is infinite list. We think that it is important to implement solvers for real problems that require rational tree support so we are going to work on it in the future, too.

\bibliographystyle{plain}
\bibliography{main}

\end{document}